\documentclass[a4paper,twoside,10pt]{article}

\input{jgrg21.sty}
\begin{document}
%
\pagestyle{fancy}
\fancyhead{}
  \fancyhead[RO,LE]{\thepage}
  \fancyhead[LO]{N. Kan}                  
  \fancyhead[RE]{Double Field Theory}
\rfoot{}
\cfoot{}
\lfoot{}
\label{P06}    
\title{%
Equations of motion in Double Field Theory: from classical particles to quantum cosmology
}
%
\author{%
  \underline{Nahomi Kan}\footnote{Email address: kan@yamaguchi-jc.ac.jp}$^{(a)}$,
  Koichiro Kobayashi\footnote{Email address: m004wa@yamaguchi-u.ac.jp}$^{(b)}$,
  and
  Kiyoshi Shiraishi\footnote{Email address:
shiraish@yamaguchi-u.ac.jp}$^{(b)}$ }
%
\address{%
  $^{(a)}$Yamaguchi Junior College, Hofu-shi, Yamaguchi 747--1232, Japan\\
  $^{(b)}$Yamaguchi University, Yamaguchi-shi, Yamaguchi 753--8512, Japan\\}
%
\abstract{
The equation of motion for a point particle in the background field of double field theory
is considered. We find that the motion is described by a geodesic flow in the doubled
geometry. Inspired by analysis on the particle motion, we propose a modified model of
quantum string cosmology, which includes two scale factors. 
The report is based on  Phys. Rev. D84 (2011) 124049 [{\tt arXiv:1108.5795}]. 
}

\section{Introduction}
Double Field Theory (DFT) \cite{P06_ref1} 
is the
theory of the massless field with a higher symmetry of spacetime including
dual coordinates. 
Through this theory, Hull, Zwiebach, and Hohm
clarified the T-duality symmetry of the massless field, and new symmetry
related to the theory. 
More recently Jeon, Lee, and Park studied the structure of DFT using
projection-compatible differential geometrical methods in
\cite{P06_ref2}. 

The present report consists of two parts. In the first part, the motion of the particle in
the background field in DFT is investigated.  We show that the geodesic in the
$2D$-dimensional doubled-spacetime cannot  be the geodesic in the $D$-dimensional
spacetime. The geodesic equation in the $D$-dimensional spacetime is found to be the
geodesic flow equation. 
In the second part, we  consider the string cosmology with a bimetric model 
inspired by the constraint method discussed in the first part.
Our method for the restriction on the metrics functions well, at least in the present
reduced model for cosmology.

\section{Review of projection-compatible approach}
Coordinates are combined
with dual coordinates to be $X^A=(\tilde{x}_a\,, \,x^{\mu})^T$,
where the suffixes $A, B,\ldots$ range over $1, 2,\ldots, 2D$,
while $\mu, \nu,\ldots$ as well as $a, b,\ldots$ range over $1, 2,\ldots, D$. 
The constant metric is assumed to be expressed as the following $2D \times 2D$ matrix,
\begin{equation}
\eta_{AB}=\left(
\begin{array}{cc}
0 & \delta^{a}{}_{\nu}\\
\delta_{\mu}{}^{b}\ & 0
\end{array}\right)\,.
\end{equation}
The suffixes are entirely raised and lowered by this constant metric.
Of course, $\eta^{AC}\eta_{CB}=\delta^{A}{}_B$ is satisfied.

The generalized metric is defined as follows:
\begin{equation}
{\cal H}_{AB}=\left(
\begin{array}{cc}
g^{ab} & ~~-g^{a\sigma}b_{\sigma\nu}\\
b_{\mu\sigma}g^{\sigma b} & ~~~g_{\mu\nu}-b_{\mu\rho}
g^{\rho\sigma}b_{\sigma\nu}
\end{array}\right)\,.
\end{equation}
Here, $g_{\mu\nu}$ and $b_{\mu\nu}$ are the metric in $D$ dimensions and the
antisymmetric tensor, respectively. It should be noted that  ${\cal H}^{AB}$ satisfes
${\cal H}^{AC}{\cal H}_{CB}=\delta^{A}{}_B$.

The following projection matrices are defined on the basis of the existence of two
kinds of metrics.
$P\equiv\frac{1}{2}(\eta+{\cal H})$,
$\bar{P}\equiv\frac{1}{2}(\eta-{\cal H})$ which satisfy
$P^2=P$, $\bar{P}^2=\bar{P}$, $P\bar{P}=\bar{P}P=0$.
From these, one can derive the identities
$P(\partial_A P)P=\bar{P}(\partial_A \bar{P})\bar{P}=0$ or $
P_{D}{}^{B}(\partial_A{\cal H}_{BC})P^{C}{}_{E}=\bar{P}_{D}{}^{B}(\partial_A{\cal
H}_{BC})\bar{P}^{C}{}_{E}=0$.

Now, the projection-compatible derivative is defined. In other words, both the metrics
are ``covariantly constant,'' {i.e.},
$\nabla_A \eta_{BC}=\nabla_A{\cal H}_{BC}=0$.
So, the covariant derivative of the projection of an arbitrary tensor coincides with
the projection of the covariant derivative of the tensor.  Jeon {et al.}\cite{P06_ref2}
found that the covariant derivatives including the following connection have the
character,
\begin{equation}
\Gamma_{ABC}\equiv
2P_{[A}{}^{D}\bar{P}_{B]}{}^{E}\partial_C
P_{DE}+2(\bar{P}_{[A}{}^{D}\bar{P}_{B]}{}^{E}-P_{[A}{}^{D}P_{B]}{}^{E})\partial_D
P_{EC}\,.
\end{equation}
They also obtained the action for the generalized metric, which was previously found
by Hohm {et al.} \cite{P06_ref1}
\[
S=\int dx d\tilde{x}\, e^{-2d}\left(\frac{1}{8}{\cal
H}^{AB}\partial_A{\cal H}^{CD}\partial_B{\cal H}_{CD}-\frac{1}{2}{\cal
H}^{AB}\partial_B{\cal H}^{CD}\partial_D{\cal H}_{AC}-2\partial_A d\partial_B{\cal
H}^{AB}+4{\cal H}^{AB}\partial_A d\partial_B d{}\right)\,,
\]
from the consideration of the projection-compatible geometrical quantities.
Here, $e^{-2d} = \sqrt{-g}\,e^{-2\phi}$ and $\phi$ is the dilaton field.  If we set
all the derivatives on the fields with respect to the dual coordinate zero
($\tilde{\partial}^a=0$), the action for the effective theory for
the zero-mode field in string theory is obtained as 
\[
S=\int dx
\sqrt{-g}\,e^{-2\phi}\left[R+4(\partial\phi)^2-\frac{1}{12}H^2\right]\,,
\]
where 
the three-form field $H=db$ is the field
strength of the Kalb-Ramond 2-form $b_{ij}$.

\section{The geodesic equation is not the equation of motion for a particle}
Next, we consider the equation of motion for a particle.
The geodesic equation is given by the
following expression
$U^\mu\nabla_\mu U^\nu=0$,
where $U^\mu\equiv \frac{dx^\mu}{ds}=\dot{x}^\mu$, $s$ being a parameter.

The corresponding equation in the projection compatible geometry of Jeon {et al.}
is considered to be 
\begin{equation}
U^A\nabla_A U^B=U^A(\partial_A U^B+\Gamma_A{}^B{}_CU^C)=0\,,
\end{equation}
where $U^A=(\tilde{U}_a, U^{\mu})^T=\frac{d{X}^A}{ds}$.
From the project space ansatz $\bar{P}U=0$, we are forced to use
$\tilde{U}_a=g_{a\nu}U^\nu$. Moreover, we set $\tilde{\partial}g=0$ as in the
interpretation of DFT. Now, we find that the above equation reads
\begin{equation}
U^\mu\partial_\mu U^\nu+\frac{1}{2}g^{\nu\mu}(\partial_{\rho}g_{\mu\sigma}+
\partial_{\sigma}g_{\mu\rho})U^\rho U^\sigma=0\,.
\end{equation}
It is obviously different from the usual geodesic equation in general relativity
(or differential geometry).
In general, it is understood that the usage of the projection has a problem.

\section{Projection and geodesic flow}
The following Lagrangian is adopted, and the mechanics derived from it are
considered:
\begin{equation}
L=\frac{1}{2}{\cal H}_{AB}\dot{X}^A\dot{X}^B+\lambda^A\bar{P}_{AB}\dot{X}^B\,.
\end{equation}
Here, $\lambda^A$ is an undecided multiplier. 
The Euler-Lagrange equation leads to the constraint $\bar{P}\dot{X}=0$.

We find that the Hamiltonian is defined as
\begin{equation}
H=
\frac{1}{2}{\cal
H}^{AB}(p_A-\lambda^C\bar{P}_{CA}) (p_B-\lambda^D\bar{P}_{DB})\,.
\end{equation}
The multiplier can be determined from the Hamilton equation as 
$\lambda_A=p_A+P_{AB}M^B$,
where $M^B$ is an arbitrary vector. 
When this is substituted into the above Hamiltonian, we obtain a new
Hamiltonian 
\begin{equation}
H_{\star}=\frac{1}{2}P^{AB}p_A p_B\,.
\end{equation}
Using the new Hamiltonian, we obtain
\begin{equation}
\dot{X}^A=\frac{\partial H_\star}{\partial p_A}=P^{AB}p_B\,,\qquad
\dot{p}_A=-\frac{\partial H_\star}{\partial X^A}=-\frac{1}{2}\partial_A
P^{BC}p_B p_C=-\frac{1}{4}\partial_A
{\cal H}^{BC}p_B p_C\,.
\end{equation}
These equations describe the geodesic flow in the system.  The combined equation
is found to be
\begin{equation}
\ddot{X}^A=\dot{X}^C(\partial_CP^{AB})p_B-\frac{1}{2}
P^{AB}\partial_BP^{CD}p_Cp_D\,.
\end{equation}

Now, let us take the condition $\tilde{\partial}^a=0$ for the correspondence with
DFT.   
If we consider $\tilde{p}^a=0$, we obtain
 $\ddot{x}^\mu+\left\{{}_{\rho\sigma}^{\,\mu\,}\right\}
\dot{x}^\rho\dot{x}^\sigma=0$, the geodesic equation in a usual $D$ dimensional
spacetime.  
We have obtained the geodesic equation in the $D$-dimensional spacetime from the
geodesic flow in the $2D$-dimensional space described by the generalized metric with
natural assumptions.

\section{A simple bi-metric model}
We apply a similar method to a modified model for
cosmology, which is related to the string cosmology \cite{P06_ref3}. In the model here, we
consider two metrics, $g$ and $\tilde{g}$. Though our model describes a bi-metric theory,
the degree of freedom is to be mildly restricted. 
For simplicity of the discussion, we consider $b_{\mu\nu}=0$.
The cosmological action we consider is
\begin{equation}
S= -\frac{\lambda_s}{2}\int d\tau\left[\frac{1}{8}
{\rm Tr}(M'\eta M'\eta)+{\Phi'}^2+ e^{-2\Phi}V\right]\,,\qquad {\rm with}\quad
M_{AB}\equiv\left(
\begin{array}{cc}
\tilde{G} & 1\\
1 & G
\end{array}\right)\,,
\end{equation}
where $G$ and $\tilde{G}$ are the spatial parts of the metrics
and $\Phi \equiv 2d$.
Here, we add the constant potential $V$ to the Lagrangian and 
$\lambda_s$ is the constant that represents the scale of string theory
\cite{P06_ref3}. 
The prime denotes differentiation with respect to $\tau$.

We now define the ``pseudo''-projection matrices
$P=\frac{\eta+M}{2}$, $\bar{P}=\frac{\eta-M}{2}$
and we wish to enforce
$P{M}'P=\bar{P}{M}'\bar{P}=0$ using some constraints.
Now, the Lagrangian $L_\Lambda$ with the constraint term is
\begin{equation}
L_\Lambda=\frac{\lambda_s}{2}\left[-\frac{1}{8}
{M'}^{AB}{M'}_{AB}+\bar{\Lambda}_{AB}\bar{P}^{AC}{M'}_{CD}\bar{P}^{DB}
+\Lambda_{AB}{P}^{AC}{M'}_{CD}{P}^{DB}-{\Phi'}^2-e^{-2\Phi}V\right]\,.
\end{equation}

The Hamiltonian of the system becomes
\begin{equation}
H_\Lambda=-\frac{4}{\lambda_s}
\left[\Pi^{AB}-\frac{\lambda_s}{2}\left(\bar{P}^{AC}\bar{\Lambda}_{CD}\bar{P}^{DB}
+{P}^{AC}\Lambda_{CD}{P}^{DB}\right)\right]^2-\frac{1}{2\lambda_s}\Pi_{\Phi}^2
+\frac{\lambda_s}{2}e^{-2\Phi}V\,,
\end{equation}
where the conjugate momentum for $M_{AB}$ and $\Phi$ are represented
by $\Pi^{AB}$ and $\Pi_\Phi$, respectively. 
We consider simplification by using the assumed relation,
$P^2\simeq P$ and $\bar{P}^2\simeq\bar{P}$.
The symbol $\simeq$ is used to indicate this assumed approximation
adopted by us.
Finally, we obtain the
Hamiltonian
\begin{equation}
H_*\equiv -\frac{8}{\lambda_s}
\Pi^{AB}{P}_{BC}\Pi^{CD}\bar{P}_{DA}-\frac{1}{2\lambda_s}\Pi_{\Phi}^2
+\frac{\lambda_s}{2}e^{-2\Phi}V\,.
\end{equation}

\section{``Minisuperspace'' version of the bi-metric model}
Next, we examine the previous procedure of modification in
the minisuperspace model.
We suppose
\begin{equation}
M_{AB}=\left(
\begin{array}{cc}
\tilde{A}(\tau)\delta^{ab} & 0\\
0 & A(\tau)\delta_{\mu\nu}
\end{array}\right)\,.
\end{equation}

The Hamiltonian for the minisuperspace version of our modified model is found to be
\begin{equation}
H_*=-\frac{2}{\lambda_s D}\,(\pi\tilde{\pi}+\tilde{\pi}\pi-A\pi
A\pi-\tilde{A}\tilde{\pi}\tilde{A}\tilde{\pi})-\frac{1}{2\lambda_s}\Pi_{\Phi}^2
+\frac{\lambda_s}{2}e^{-2\Phi}V\,.
\end{equation}

A special solution can be found for the Hamilton equations. 
The solution is
\begin{equation}
\tilde{A}(\tau)=\frac{1}{A_0}\exp\left[-\frac{2}{\sqrt{D}}C(\tau-\tau_0)\right]\,,\quad
{A}(\tau)={A_0}\exp\left[\frac{2}{\sqrt{D}}C(\tau-\tau_0)\right]+\delta\,,
\end{equation}
where $A_0$, $\tau_0$, and $\delta$ are constants.
The similarity to the known string cosmological solution \cite{P06_ref3} is obvious, up to
the possible constant deviation $\delta$ in $A$.
For the solution, we find that $A\tilde{A}\rightarrow 1$ when $\tau\rightarrow
+\infty$.

\section{Quantum cosmology of the bi-metric model}
Quantum cosmological treatment of the string cosmology has been widely studied
\cite{P06_ref3}. In our model, the minisuperspace Wheeler-DeWitt
equation is obtained as
\begin{equation}
\left[\frac{2}{\lambda_s
D}\,\left(2\frac{\partial}{\partial A}\frac{\partial}{\partial
\tilde{A}}-A\frac{\partial}{\partial A}
A\frac{\partial}{\partial
A}-\tilde{A}\frac{\partial}{\partial
\tilde{A}}\tilde{A}\frac{\partial}{\partial
\tilde{A}}\right)+\frac{1}{2\lambda_s}
\frac{\partial^2}{\partial\Phi^2}
+\frac{\lambda_s}{2}e^{-2\Phi}V\right]\Psi=0\,,
\end{equation}
where $\Psi$ is the wave function of the universe.
To simplify the description of the system, we use the following variables:
$x=\frac{\sqrt{D}}{4}\ln A/\tilde{A}$, $y=\frac{\sqrt{D}}{4}\ln A\tilde{A}$.
Up to the ordering, we have 
\begin{equation}
\left[-\frac{1+e^{-({4}/{\sqrt{D}})y}}{2}\frac{\partial^2}{\partial
x^2}-\frac{\partial}{\partial
y}\frac{1-e^{-({4}/{\sqrt{D}})y}}{2}\frac{\partial}{\partial y}+
\frac{\partial^2}{\partial\Phi^2}
+{\lambda_s^2}e^{-2\Phi}V\right]\Psi=0\,.
\end{equation}

If we assume a solution of the form,
$\Psi(x,y,\Phi)=X_k(x)Y_{kK}(y)Z_K(\Phi)$, we find
the non singular real solution for $Y_{kK}(y)$ at $y=0$ as follows:
\begin{equation}
Y_{kK}(y)=e^{-\sqrt{k^2-2K^2}y}F\left(
{
\frac{1+\sqrt{1-k^2}+\sqrt{k^2-2K^2}}{2},
\frac{1-\sqrt{1-k^2}+\sqrt{k^2-2K^2}}{2},1;1-e^{-2y}
}
\right)\,,
\end{equation}
where $F(\alpha,\beta,\gamma; z)$ is the Gauss' hypergeometric function.

\begin{figure*}[h]
\centering
\includegraphics[keepaspectratio=true,width=7cm]
{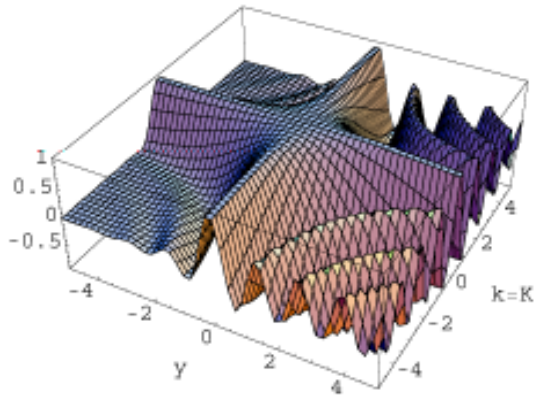}
\caption{%
3D-plot of $Y_{kk}(y)$.
}
\label{P06_fig1.eps}
\end{figure*}

If $K=\pm k$, $Y_{k\,\pm k}(y)$ has a maximum at $y=0$ (see Figure~\ref{P06_fig1.eps}).
When we construct a wave packet for the cosmological wave function \cite{P06_ref3}, the
peak of this wave packet in terms of parameter $y$ is naturally located at $y=0$. Thus,
the approximate scale factor duality $A\tilde{A}\simeq 1$ is expected even at the
``beginning'' of the quantum universe. The detailed investigation on the behavior of the
universe is left for future research.

\end{document}